\documentstyle[twoside,fleqn,espcrc2]{article}

\newcommand{\AmS}{{\protect\the\textfont2
  A\kern-.1667em\lower.5ex\hbox{M}\kern-.125emS}}

\title{M-Theory, Torons and Confinement}

\author{C. G\'{o}mez and R. Hern\'{a}ndez\address{Instituto de Matem\'{a}ticas y F\'{\i}sica Fundamental,
CSIC, Serrano 123, 28006 Madrid, Spain}%
        \thanks{This research is partially supported by under grant AEN $96$-$1655$, and 
by E. C. grant FMRX-CT $960012$.}}

\begin{document}

\begin{abstract}
We study the decompactification limit of M-theory superpotentials for
$N\!=\!1$ four dimensional supersymmetric gauge theories. These superpotentials
can be interpreted as generated by toron configurations. The connection
with the confinement picture in the maximal abelian gauge is discussed.
\end{abstract}

\maketitle

%\section{}

Polyakov's compact electrodynamics in three dimensions \cite{P} constitutes
a pardigmatic model for electric confinement. In the late seventies,
Callan, Dashen and Gross \cite{CDG} were tryig to extend the peculiar
features of this model, namely the Coulomb interaction between three
dimensional instantons, to four dimensions. In order to achieve this goal,
they suggested a confinement mechanism based on topologically
charged objects, with a fractional charge, the merons \cite{F}. Using M-theory
compactifications on an $SU(4)$ holonomy Calabi-Yau fourfold $X$, most
of Polyakov's dynamics can be now derived. In particular,
the instanton generated $N\!=\!2$ superpotential in three dimensions
\cite{AHW},
\begin{equation}
W=exp (-S_0-i \phi_D),
\end{equation}
with $S_0$ the classical instanton action, and $\phi_D$ the dual photon
field in three dimensions, can be derived using M-theory instantons \cite{Wsp},
defined by wrapping the M-theory $5$-brane on a $6$-cycle $D$ of $X$,
with holomorphic aritmetic genus $\chi(D)=1$. Provided the fourfold $X$ is
elliptically fibered,
\begin{equation}
E \longrightarrow X \stackrel{\Pi}{\longrightarrow} B,
\end{equation}
and $D$ is a vertical instanton \cite{Wsp}, i. e., $\pi(D) \subset B$, then
its effect survives the four dimensional $\epsilon \rightarrow 0$ limit,
where $\epsilon$ is the K\"{a}hler class of the elliptic fiber $E$, so that
we can get, in the decompactification limit, instanton contributions
to the $N\!=\!1$ four dimensional superpotentials \cite{KV,BJPSV}.

\vspace{2 mm}

{\bf Super Yang-Mills Instantons.} It is well known \cite{NSVZ,A}
that for pure $SU(N_C)$ $N\!=\!1$ supersymmetric Yang-Mills,
instantons generate gluino condensates of the type
\begin{equation}
< \prod_{i=1}^{N_C} \lambda \lambda (x_i) >= \hbox {constant}
\Lambda^{3N_C},
\label{eq:3}
\end{equation}
with $\Lambda^{3}=\mu^3 \hbox {exp} \frac {-8 \pi^2}{N_C g^2}$.
Correlators of type (\ref{eq:3}) are usually interpreted in terms
of cluster decomposition, as products of $<\lambda \lambda>$
condensates,
\begin{equation}
< \lambda \lambda >= \hbox {constant } e^{2 \pi i e/N_C}
\Lambda^{3},
\label{eq:4}
\end{equation}
with $e=0, \ldots, N_C-1$, corresponding to the $N_C$ different
vacua predicted by the $\hbox {tr} (-1)^{F}$ computation
\cite{Wind}. The independence of (\ref{eq:3}) on the positions
$x_i$ is due to the integration over the instanton size required
by supersymmetry. In particular, the correlator $<\lambda \lambda
(x_1) \lambda \lambda (x_2)>$ for $SU(2)$ is saturated by
instantons with size $\rho \sim |x_2 -x_1|$ \cite{NSVZ}. It is
clear from (\ref{eq:3}) that instantons do not generate a
superpotential for $N\!=\!1$ pure super Yang-Mills; however, it
is in principle possible to define an effective superpotential
\cite{KS} with a set of $N_C$ minima characterized by the expectation
values given by (\ref{eq:4}). On the other hand, we can use the exact
solution for $N\!=\!2$ pure Yang-Mills \cite{SW} and softly break
to $N\!=\!1$ through the addition of a mass term. In this case, and for 
$SU(2)$ gauge group, the following superpotential for the monopole hypermultiplet 
is derived \cite{SW}:
\begin{equation}
W= a_D M \tilde{M} + m u(a_D),
\label{eq:5}
\end{equation}
  
The two minima of (\ref{eq:5}) are now parametrized by a vacuum
expectation value of the monopole field, which can be
interpreted, through the dual Higgs mechanism, as the order
parameter for confinement. It is, of course, tempting to identify
the two minima of (\ref{eq:5}) with the ones described by
(\ref{eq:4}) in the $SU(2)$ case. This is an interesting
dynamical issue, as the condensates (\ref{eq:4}) are in principle
instanton generated, while (\ref{eq:5}) is the effective
potential for the monopoles, that from the $N\!=\!1$ point of
view, we should interpret as defining `t Hooft's abelian
projection \cite{tHap} for pure $N\!=\!1$ supersymmetric Yang-Mills
theory. Of course, the interplay between confinement and the
derivation of the $<\lambda \lambda>$ condensates from the
instanton amplitude (\ref{eq:3}) is contained in the cluster
decomposition argument.
  
\vspace{2 mm}
  
{\bf Torons.} A different way to derive the condensate
(\ref{eq:4}) using `t Hooft's torons \cite{tHcmp} was suggested
some time ago in reference \cite{CG} \footnote{See also references 
\cite{Z} and \cite{LS}.}. Toron configuration for
$SU(N_C)$ gauge theories are defined using twisted boundary
conditions \cite{tHtw} on a $T^4$ box of size $L$. Denoting by $n_{ij}$ the
twist in the $(i,j)$-plane, the toron of Pontryagin number $\frac
{1}{N_C}$ is defined by $n_{12}=-1$, $n_{34}=1$. With the
definitions $m_i \equiv \epsilon_{ijk}n_{jk}$ and $k_i \equiv
n_{i4}$, the toron configuration can be interpreted as a
tunneling between the states $|m_3=1, k_3>$ and $|m_3,k_3+1>$. In
terms of the invariant states
\begin{equation}
|e,m> \equiv \frac {1}{N_C^3} \sum_{k} e^{2 \pi i k e/N_C} |k,m>,
\label{eq:6}
\end{equation}
the toron amplitude for $<\lambda \lambda>$ in the $L \rightarrow \infty$
limit is given by \cite{CG}
\begin{eqnarray}
<e,m_3|\lambda \lambda(x)|e,m_3> & = &   \nonumber \\ 
\frac {1}{N_C} \sum_{k_3}
<k_3+1,m_3|\lambda \lambda(x) |k_3,m_3=1> &  = &   \nonumber \\  
\Lambda^3 e^{2 \pi e/N_C}, &   & 
\label{eq:7}
\end{eqnarray}
where now $e$ represents `t Hooft's electric flux in the
$3$-direction, and can take values $e=0,1,\ldots,N_C-1$, in
agreement with result (\ref{eq:4}). Notice that the toron action
is given by 
\begin{equation}
S_{toron}(e,\theta) = \frac {8 \pi^2}{N_C g^2} + \frac {2 \pi i
e}{N_C} + \frac {i \theta}{N_C},
\label{eq:8}
\end{equation}
in contrast to the instanton action $\frac {8 \pi^2}{g^2} + i
\theta$ \footnote{Notice that the toron action depends on the
$\theta$-parameter, and on the electric flux $e$. In fact, it is
the electric flux $e$ the one that is going to play a similar
dynamical role to the dual photon in the $2+1$ dimensional
example.}.  
  
\vspace{2 mm}

{\bf Toron Superpotential.} Let us now for a moment assume that torons are the right
configuration contributing to the $\lambda \lambda$-condensates.
If such is the case, taking into account that they have only two
fermionic zero modes, we can in principle use this configurations
to generate a superpotential for $N\!=\!1$, in such a way
that their minima reproduce the correct values for the $\lambda
\lambda$-condensates \cite{GH}. In the same way as for the $2+1$ $N\!=\!2$
example of reference \cite{AHW}, we will need some information on
how to define a toron dilute gas, and how to define toron
correlations. Let us imagine that torons are
``instanton-pieces'', and that we divide the $SU(N_C)$ instanton
into $N_C$ torons. Morever, we formally leave the electric flux
$e$ in (\ref{eq:8}) as a free parameter; then, the
superpotential, in dilute gas aproximation, for these
$N_C$-torons would be given by
\begin{equation}
W = \sum _{i=1}^{N_C} \hbox{exp} (- S_{toron}(e_i)).
\label{eq:9}
\end{equation}

Now, as we impose the constraint that the $N_C$-torons are the
pieces of an ordinary $SU(N_C)$ instanton, and using (\ref{eq:8}) we obtain
\begin{eqnarray}
W= \hbox {exp} \frac {-8 \pi^2}{N_C g^2} \sum_{i=1}^{N_C-1} \hbox{exp} 
\left( \frac {-2 \pi i e_i}{N_C} \right) & + & \nonumber \\  \hbox{exp} \frac {-8 \pi^2}{g^2} \hbox {exp}
\sum_{i=1}^{N_C-1} \left( \frac {8 \pi^2}{g^2 N_C} + \frac {2 \pi
i e_i}{N_C} \right) &  & .
\label{eq:10}
\end{eqnarray}
  
Notice that in (\ref{eq:10}) we have used $\theta=0$; however, no
difference arises if $\theta$ is left free and common to torons
and instantons. It is obvious that (\ref{eq:10}), as a
superpotential for $e_i$'s, possesses a set of $N_C$ minima for
$e=0, \ldots, N_C-1$. It is important to stress that instantons
are blind to the value of $e$, and that we fix its value at the
minima of (\ref{eq:10}) only after impossing the constraint that
the ordinary instanton is a ``dilute gas'' of $N_C$ torons. In
fact, the previous discussion does not imply any probelm with the
standard periodicity of the $\theta$-angle.
  
\vspace{2 mm}

{\bf M-Theory Model.} Let us now try to see how much we can
learn about the origin of $\lambda \lambda$-condensates in
$N\!=\!1$ using M-theory techniques. A model for 
$N\!=\!1$ supersymmetric Yang-Mills theory in four dimensions
can be engineered using F-theory compactifications \cite{KV} on an
elliptically fibered Calabi-Yau fourfold with $SU(4)$ holonomy. Let $C$
be a divisor in $B$ satisfying
\begin{equation}
h^{1,0}=h^{2,0}=0,
\label{eq:11}
\end{equation}
and such that the elliptic fiber $E$ on $C$ is singular, of ADE type in
Kodaira's classification; a partially wrapped $7$-brane on $C$ will
define an $N\!=\!1$ ADE gauge theory in four dimensions. From now
on, we will consider the simpler case of $A_{N_C-1}$
singularities, corresponding to $SU(N_C)$ $N\!=\!1$
supersymmetric Yang-Mills in four dimensions.
  
From the point of view of M-theory, we can think of a
compactification on an elliptically fibered Calabi-Yau, and take
the limit $\hbox{Vol}(E) \rightarrow 0$. The resulting
theory is four dimensional, with the extra dimension obtained by
interpreting membranes wrapped on $E$ as Kaluza-Klein states \cite{SS}. The
singular fiber $E$ of type $A_{N_C-1}$ has Euler characteristic 
\begin{equation}
\chi(E) = N_C,
\label{eq:12}
\end{equation}
corresponding to a set of $N_C$ irreducible components $E_i$,
with $E_i \cdot E_i=-2$, and intersection matrix given by the
corresponding affine Dynkin diagram, $\hat{A}_{N_C-1}$ \footnote{In order to obtain the Euler 
characteristic $2$ should be assigned to each point, and $-1$ to each link.}. 
  
As was shown in \cite{KV}, we can use the irreducible components
$E_i$ to define M-theory instantons with holomorphic genus equal
one, with the corresponding divisor obtained by fibering $E_i$
over $C$. Each of these instantons would be characterized by the
action 
\begin{equation}
S=V(C)V(E_i)=\frac {8 \pi^2}{g_4^2} \frac {\hbox {Vol}(E_i)}{\hbox {Vol}(E)}.
\label{eq:14}
\end{equation}
Using the Kaluza-Klein \cite{SS} interpretation of membranes
wrapped on $E$, we can write (\ref{eq:14}) as $\frac {R}{g_4^2}
\hbox {Vol} (E_i)=\frac {1}{g_3^2} \hbox {Vol}(E_i)$, and to
interpret $\hbox {Vol}(E_i)$ as the parameters of the Coulomb
branch of the $2+1$ dimensional theory. Notice that the imaginary
part of $\hbox {Vol}(E_i)$ represents the Wilson loop in the
internal direction of effective radius $R$. The superpotential
generated by these M-theory instantons is given by
\begin{equation}
W = \sum _{i=1}^{N_C} \hbox{exp} (- V(C)V(E_i)),
\label{eq:15}
\end{equation}
with the constraint
\begin{equation}
\sum_{i=1}^{N_C} E_i =E,
\label{eq:16}
\end{equation}
reflecting the fact that the $E_i$ are the $N_C$ components of
the singular fiber $E$. In the decompactification limit $\hbox
{Vol}(E) \rightarrow 0$, the superpotential (\ref{eq:15}), with
(\ref{eq:16}), becomes
\begin{eqnarray}
W=\sum_{i=1}^{N_C-1} \hbox{exp} (-V(C)V(E_i)) & + & \nonumber \\
\gamma \hbox{exp} \sum_{i=1}^{N_C-1} (V(C)V(E_i)), &  &
\label{eq:17}
\end{eqnarray}
with $\gamma = \hbox{exp} \left( \frac {-8 \pi^2}{g_3^2 R}
\right)$, and the minima corresponding to the toron action $\frac
{8 \pi^2}{N_Cg_4^2} + \frac {2 \pi i e}{N_C}$, with $e=0, \ldots,
N_C-1$.
  
Thus, the partition of the M-theory divisor $E$ into irreducible
components $E_i$ is the algebraic geometrical analog of the
partition of the instanton into torons. Moreover, in the M-theory
approach the `t Hooft electric flux $e$, that we have left free
in (\ref{eq:10}), becomes the blow up coordinate of the singular
fiber. We minimize a configuration relative to these coordinates,
constrained by the decompactification condition $Vol(E)=0$, in a
similar way as when we constrained the values of electric flux
for torons by impossing that the set of $N_C$ torons define an
instanton.
  
\vspace{2 mm}
  
{\bf Four Dimensions From Three Dimensions.} A different 
approach for obtaining the superpotential (\ref{eq:17}) for
$SU(2)$ was presented in reference \cite{SW3d}. This approach uses the
Atiyah-Hitchin hyperk\"{a}hler manifold \cite{AH} for three dimensional
$N\!=\!4$ supersymmetric Yang-Mills theory. The derivation starts from the
superpotential for the $N\!=\!2$ Seiberg-Witten solution \cite{SW},
softly broken to $N\!=\!1$,
\begin{equation}
W = \lambda (y^2-x^2(x-u)-x) + m u.
\label{eq:12b}
\end{equation}
The corresponding Atiyah-Hitchin manifold is recovered upon compactification
on $S^1$, through the definition $\frac {1}{g_3^2}=\frac {R}{g_4^2}$, in the
three dimensional limit, as $m \rightarrow 0$. This can be achieved
using the following change of variables \cite{SW3d}: $v=x-u$, $x=\gamma \tilde{x}$,
$y =\gamma \tilde{y}$, with $\gamma = \hbox{exp} \left( \frac {-8 \pi^2}{g_3^2R}
\right)$. The superpotential (\ref{eq:12b}) now becomes
\begin{equation}
W = m \left( \frac {1}{\tilde{x}} + \gamma \tilde{x} \right),
\label{eq:13b}
\end{equation}
which coincides with (\ref{eq:10}) for $\frac {\tilde{x}}{m} \equiv \hbox{exp}(V_i)$.
  
\vspace{2 mm}
  
{\bf Abelian Projection and Confinement.}  The crucial step in
our previous discussion was to interpret the blow up parameters
$V(E_i)$ in the $\hbox{Vol}(E_i) \rightarrow 0$ limit as the
electric fluxes that appear in the toron action (\ref{eq:8}). A possible way
to interpret these coefficients in the $\hbox{Vol}(E_i)
\rightarrow 0$ limit would be as defining the abelian projection
gauge \cite{tHap}. In the abelian projection philosophy, the
gauge is determined dynamically either by the vacuum expectation
value of the monopole field or, in electric variables, by the
type of gauge configurations that contribute to confinement. If
our interpretation of the superpotential (\ref{eq:17}) in term of torons is
correct, it seems that confinement should be intimatelly related
with the effective breaking of instantons into toron pieces. This
effective breaking of the instanton, with a very close look to
the toron liquid picture \cite{GMM}, is very natural from the M-theory point
of view where algebraic geometry dictates the representation of
the $A_{N_C-1}$ singular fiber $E$ in terms of $N_C$ irreducible
components.
  
It would be very interesting to consider instantons in the
abelian projection as sets of $N_C$ torons. An important role in
that direction should be played by the size of the instanton. In
fact, for the instanton in the singular gauge,
\begin{equation}
A_{\mu}^{a} = 2 \frac {\rho^2}{x^2+ \rho^2} \bar{\eta}_{a \mu \nu}
\frac {x_{\nu}}{x^2},
\label{eq:18}
\end{equation}
the maximal abelian gauge is obtained by minimization of the
functional \cite{KSW,BOT,lattice}
\begin{equation}
G = \frac  {1}{4} \int d^4x ( A_{\mu}^{1}(x)^2 + A_{\mu}^{2}(x)^2
),
\label{eq:19}
\end{equation}
which takes the value $4 \pi \rho^2$.   
  
\vspace{2 mm}
  
In summary, we conclude that a fractionalization of instantons is
crucial for the confinement dynamics in $N\!=\!1$ supersymmetric
Yang-Mills theory, and that the M-theory description of
instantons is the appropiate framework to understand the
underlying dynamics.


\begin{thebibliography}{9}

\bibitem{P} A. M. Polyakov, ``Quark Confinement and the Topology of Gauge 
Groups'', Nucl. Phys. {\bf B120} (1977), 429.

\bibitem{CDG} C. G. Callan, R. Dashen and D. J. Gross, ``Towards
a Theory of the Strong Interactions'', Phys. Rev. {\bf D17} (1978), 2717.

\bibitem{F} V. De Alfaro, S. Fubini and G. Furlon, ``A New
Classical Solution of the Yang-Mills Field Equations'', Phys. Lett.
{\bf B65} (1976), 164.

\bibitem{AHW} I. Affleck, J. A. Harvey and E. Witten, ``Instantons and 
Supersymmetry Breaking in $2+1$ Dimensions'', Nucl. Phys. {\bf B206} (1982), 413.

\bibitem{Wsp} E. Witten, ``Non-Perturbative Superpotentials in String Theory'',  
Nucl. Phys. {\bf B474} (1996), 343.

\bibitem{KV} S. Katz and C. Vafa, ``Geometric Engineering of $N\!=\!1$ 
Quantum Field Theories'', hep-th/9611090.

\bibitem{BJPSV} M. Bershadsky, A. Johansen, T. Pantev, V. Sadov and C. Vafa, 
``F-Theory, Geometric Engineering and $N\!=\!1$ Dualities'', hep-th/9612052.

\bibitem{NSVZ} V. Novikov, M. Shifman, A. Vainshtein and V. Zakarov, Nucl. Phys. {\bf B229} (1983), 
381; Nucl. Phys. {\bf B229} (1983), 394; Nucl. Phys. {\bf B229} (1983), 407.

\bibitem{A} D. Amati, K. Konishi, Y. Meurice G. C. Rossi and G. Veneziano, 
``Non Perturbative Aspects in Supersymmetric Gauge Theories'', Phys. Rep. 
{\bf 162} (1988), 169.

\bibitem{Wind} E. Witten, ``Constrains on Supersymmetry Breaking'', 
Nucl. Phys. {\bf B202} (1982), 253.

\bibitem{KS} A. Kovner and M. Shifman, ``Chirally Symmetric Phase of
Supersymmetric Gluodynamics'', {\bf hep-th/9702174}.

\bibitem{SW} N. Seiberg and E. Witten, ``Electric-Magnetic Duality, Monopole
Condensation and Confinement in $N\!=\!2$ Supersymmetric Yang-Mills Theory'',
Nucl. Phys. {\bf B426} (1994), 19.

\bibitem{tHap} G. `t Hooft, ``Topology of the Gauge Condition and
New Confinement Phases in Non Abelian Gauge Theories'', Nucl. Phys. {\bf B190}
(1981), 455.

\bibitem{tHcmp} G. `t Hooft, ``Some twisted Self-Dual Solutions for the 
Yang-Mills Equations on a Hypertorus'', Commun. Math. Phys. {\bf 81} (1981), 267.

\bibitem{CG} E. Cohen and C. G\'{o}mez, ``Chiral Symmetry Breaking in 
Supersymmetric Yang-Mills'', Phys. Rev. Lett. {\bf 52} (1984), 237.

\bibitem{Z} A. R. Zhitnitsky, ``Torons, Chiral Symmetry Breaking
and the $U(1)$ Problem in the $\sigma$-Model and in Gauge
Theories'', Nucl. Phys. {\bf B340} (1990), 56.

\bibitem{LS} H. Leutwyler and A. Smilga, ``Spectrum of the
Dirac Operator and Role of Winding Number in QCD'', Phys. Rev. {\bf D46}
(1992), 5607.

\bibitem{tHtw} G. `t Hooft, ``A Property of Electric and Magnetic Charges
in Non Abelian Gauge Theories'', Nucl. Phys. {\bf B153} (1979), 141.

\bibitem{GH} C. G\'{o}mez and R. Hern\'{a}ndez, ``M and F-Theory
Instantons, $N\!=\!1$ Supersymmetry and Fractional Topological Charge'',
{\bf hep-th/9701150}.

\bibitem{SS} S. Sethi and L. Susskind, ``Rotational Invariance in
the M(atrix) Formulation of Type II$_B$ Theory'', {\bf hep-th/9702101}.

\bibitem{SW3d} N. Seiberg and E. Witten, ``Gauge Dynamics and Compactification
to Three Dimensions'', hep-th/9607163.

\bibitem{AH} M. F. Atiyah and N. Hitchin, {\em The Geometry and Dynamics 
of Magnetic Monopoles}, Princeton University Press, 1988. 

\bibitem{GMM} A. Gonz\'{a}lez-Arroyo. P. Mart\'{\i}nez and A. Montero,
Phys. Lett. {\bf B359} (1995), 159.

\bibitem{KSW} A. S. Kronfeld, G. Schierholz and U. Wiese, Nucl. Phys.
{\bf B293} (1987), 461.

\bibitem{BOT} R. C. Brower, K. N. Orgines and C. I. Tann, ``Instantons
in the Maximal Abelian Gauge'', Nucl. Phys. Proc. Suppl. {\bf B53} (1997), 488.

\bibitem{lattice} See, for instance, ``Lattice $96$'', Nucl. Phys.
{\bf B} Proc. Suppl. {\bf 53} (1997).




\end{thebibliography}
\end{document}